\documentclass{svjour3}                     
\smartqed  
\usepackage{cite}
\usepackage{graphicx}
 \journalname{}
\makeatletter
\AtBeginDocument{\@ifpackageloaded{natbib}{\ifNAT@numbers\if@filesw\immediate\write\@auxout{\string\global\string\NAT@numberstrue}\fi\fi}{}}
\makeatother
\usepackage{amstext}

\begin{document}

\title{Hawking radiation from a five-dimensional Lovelock black hole}


\author{Mahamat Saleh         \and
       Bouetou Bouetou Thomas \and
         Timoleon Crepin Kofane 
}


\institute{Mahamat Saleh \and
           Bouetou Bouetou Thomas \and Timoleon Crepin Kofane \at
              Department of Physics, Faculty of Science, University of Yaound\'e I, P.O. Box 812, Yaound\'e, Cameroon \\
           \and
           Mahamat Saleh
             \at
              Department of Physics, Higher Teachers' Training College, University of Maroua, P.O. Box 55, Maroua, Cameroon \\\email{mahsaleh2000@yahoo.fr}
              \and
              Bouetou Bouetou Thomas
           \at
              National Advanced School of Engineering, University of Yaound\'e I, P.O. Box 8390, Yaound\'e, Cameroon \\\email{tbouetou@yahoo.fr}
              \and
              Kofane Timoleon Crepin
              \at
              The Max Planck Institute for the Physics of Complex Systems,
N\"othnitzer Strasse 38, 01187 Dresden, Germany \\\email{tckofane@yahoo.com}
            }

\date{Received: date / Accepted: date}

\maketitle

\begin{abstract}
We investigate Hawking radiation from a five-dimensional Lovelock black hole  using the Hamilton-Jacobi method. The behavior of the rate of radiation is plotted for various values of  the ultraviolet correction parameter and the cosmological constant. The results show that, owing to the ultraviolet correction and the presence of dark energy represented by the cosmological constant, the black hole radiates at a slower rate in comparison to the case without ultraviolet correction or cosmological constant. Moreover, the presence of the cosmological constant makes the effect of the ultraviolet correction on the black hole radiation negligible.

\end{abstract}

%

\section{Introduction}
\label{intro}
Hawking radiation,
which is closely related to the existence of a  black hole's event horizon,
 is an important quantum phenomenon. Hawking radiation from black holes\cite{jiang, yu, zhangxu, zhao1, jiangkepeng, passaoglu, pw, 31225, 31370, haw1, haw2} is one of the most striking effects  known or at least widely agreed to arise from the combination of quantum mechanics and general relativity. As one of the most important achievements of quantum field theory in curved spacetimes, the discovery of Hawking radiation lent support
to the idea that a classical black hole could radiate a thermal spectrum of particles.
Since Hawking's original work, several derivations of Hawking radiation have been proposed in the literature.
Kraus and Wilczek\cite{31225, 31370} considered the  modification of the formulas for black hole radiation  resulting from the self-gravitation
of  the  radiation and found  that  the  particles  no  longer move  along  geodesics  and that  the  action  along  the  rays is no longer  zero
for  a  massless  particle. They concluded that the  radiation  is  no  longer thermal  but  can be  corrected in  a  definite  way that
they calculated. In 2000, Parikh and Wilczek, elaborating upon
Kraus and Wilczek's work, presented another new derivation of Hawking radiation, in which Hawking radiation is treated as
a quantum tunneling process\cite{pw}. By using this method, many studies were conducted to evaluate the black hole radiation from massless particles\cite{hezhang}, massive particles\cite{deyou1}, charged massive particles\cite{zhao, jiang, qqjiang}, and Dirac particles\cite{deyou, qiru, qi}. In 2012, Jiang and Han investigated black hole spectroscopy via adiabatic invariance by combining the black hole property of adiabaticity with the oscillating velocity of the black hole horizon obtained from the tunneling framework\cite{jianghan}. Recently, some related works have developed Jiang-Han's method to investigate the entropy spectra of different black holes\cite{r1,r2,r3,r4,r5,GAL,SNZ}. Many other works involving the Hamilton-Jacobi method to investigate Hawking radiation can be found in the literature\cite{deyou1, vanzo, ding, linyang, shao, liuchenyang,mah}.  These works highlight the fact that black holes are not exclusively absorbing;  they are also emitting radiation.

Recently, Cai \emph{et al.}\cite{caicao} investigated Hawking radiation of an apparent horizon in a Friedmann-Robertson-Walker
universe using the Hamilton-Jacobi method. Using the same method, Gohar and Saifullah\cite{gohar} investigated scalar field radiation from dilatonic black holes. In this paper, a five-dimensional Lovelock black hole is
considered to investigate Hawking radiation by including the
influence of the ultraviolet correction to the black hole.

\section{Hawking radiation of the black hole}
\label{sec:1}
The five-dimensional Lovelock black hole metric is given by\cite{matias,chenwang}
\begin{equation}\label{1}
ds^{2}=f(r)dt^{2}-f(r)^{-1}dr^{2}-r^{2}d\Omega^2_3,
\end{equation}
where $d\Omega^2_3=d\theta^{2}+\sin^{2}\theta
d\varphi^{2}+\sin^2\theta \sin^2\varphi d\psi^2$,

\begin{equation}\label{2}
    f(r)=\frac{4\alpha-4M+2r^2-\Lambda r^4/3}{4\alpha+r^2+\sqrt{r^4+\frac{4}{3}\alpha\Lambda r^4+16M\alpha}},
\end{equation}
  $M$ is the black hole mass, $\Lambda$ is the cosmological constant, and
$\alpha$ is the coupling constant of an additional term that represents the ultraviolet correction to Einstein theory.

For $\Lambda=0$, the radial function (\ref{2}) reduces to
\begin{equation}\label{02}
    f(r)=\frac{4\alpha-4M+2r^2}{4\alpha+r^2+\sqrt{r^4+16M\alpha}},
\end{equation}
and the horizon radius is
\begin{equation}\label{03}
    r_H=\sqrt{2(M-\alpha)}.
\end{equation}

When $\Lambda\neq0$, the horizon radii, $r_+$ and $r_-$, for this background are given by
\begin{equation}\label{eq4}
    r_+=\sqrt{\frac{3}{\Lambda}(1+\sqrt{1-4\Lambda(M-\alpha)/3})}
\end{equation}
and
\begin{equation}\label{4}
    r_-=\sqrt{\frac{3}{\Lambda}(1-\sqrt{1-4\Lambda(M-\alpha)/3})}.
\end{equation}
We can then convert the metric into the following form:
\begin{equation}\label{5}
    f(r)=\frac{-\frac{\Lambda}{3}(r^2-r_+^2)(r^2-r_-^2)}{4\alpha+r^2+\sqrt{r^4+\frac{4}{3}\alpha\Lambda r^4+16M\alpha}}
\end{equation}

The Hamilton-Jacobi method is an alternate method for calculating black hole tunneling that
makes use of the Hamilton-Jacobi equation as an ansatz\cite{agheben}. This method
is based on the work of
Padmanabhan and his collaborators\cite{pad1, pad2, pad3}. In general, the method involves using the WKB approximation to solve a wave equation. The simplest
case to model is that of scalar particles, which therefore involves applying the WKB
approximation to the Klein-Gordon equation.  The result, to the lowest order
of the WKB approximation, is a differential equation that can be solved by substituting a suitable ansatz. The ansatz is chosen by using the symmetries of
the spacetime to assume separability. After substituting a suitable ansatz, the
resulting equation can be solved by integrating along the classically forbidden
trajectory, which starts inside the horizon and finishes at the outside observer
(usually at infinity). Because this trajectory is classically forbidden, the equation
will have a simple pole located at the horizon. Consequently, the
method of complex path analysis must be applied to  deflect the path around the pole.

Scalar particles under a gravitational background obey the Klein-Gordon
equation. For a scalar particle moving in spacetime, the radiated particle obeys the Klein-Gordon equation for a scalar field $\phi$:
\begin{equation}\label{hj1}
    g^{\mu\nu}\partial_\mu\partial_\nu\phi-\frac{m^2}{\hbar^2}\phi=0.
\end{equation}
Applying the WKB approximation by assuming an ansatz of the form
\begin{equation}\label{hj2}
    \phi(t,r,\theta,\varphi)=\exp[\frac{i}{\hbar}I(t,r,\theta,\varphi)+I_1(t,r,\theta,\varphi)+O(\hbar)],
\end{equation}
where $I$ and $I_1$ are the components of the action approximated at the zeroth and first order, respectively,
and then inserting this back  into the Klein-Gordon equation  will result in the
Hamilton-Jacobi equation to the lowest order in $\hbar$:
\begin{equation}\label{hj3}
    g^{\mu\nu}\partial_\mu I\partial_\nu I+m^2=0.
\end{equation}
For the Hamilton-Jacobi ansatz,  the classically forbidden trajectory
from inside to outside the horizon is given by\cite{agheben}

\begin{equation}\label{hj4}
    \Gamma\propto\exp(-2\textrm{Im} I).
\end{equation}
For the five-dimensional Lovelock black hole metric, the Hamilton-Jacobi equation is explicitly
\begin{equation}\label{hj5}
\begin{array}{lll}
    -f^{-1}(r)(\partial_tI)^2+f(r)(\partial_rI)^2+\frac{1}{r^2}(\partial_\theta I)^2&&\\
    +\frac{1}{r^2\sin^2\theta}(\partial_\varphi I)^2
    +\frac{1}{r^2\sin^2\theta \sin^2\varphi}(\partial_\psi I)^2+m^2=0.
    \end{array}
\end{equation}
Considering the symmetry of the black hole metric, we perform the following
separation of variables for the action $I$:
\begin{equation}\label{hj6}
    I(t,r,\theta,\varphi)=-\omega t+W(r)+J(\theta,\varphi).
\end{equation}
As a consequence, we have
\begin{equation}\label{hj7}
    \partial_t I=-\omega; \ \ \partial_r I=W'(r); \ \ \partial_\theta I=J_\theta; \ \ \partial_\varphi I=J_\varphi,
\end{equation}where
$J_\theta$ and $J_\varphi$ are constants.
Since $\partial_t$ is the time-like
killing vector for this coordinate system, $\omega$ is the energy of the particle as
detected by an observer at infinity.

 Having these expressions, we can transform Eq. (\ref{hj5})  to
\begin{equation}\label{hj8}
    -\omega^2 f^{-1}(r)+f(r)(W'(r))^2+r^{-2}(J_\theta)^2+r^{-2}\sin^{-2}\theta(J_\varphi)^2+m^2=0.
\end{equation}
Solving for $W(r)$ yields
\begin{equation}\label{hj9}
    W_\pm(r)=\pm\int{\frac{dr}{f(r)}\sqrt{\omega^2-f(r)(m^2+r^{-2}J^2_\theta+r^{-2}\sin^{-2}\theta J^2_\varphi)}}
\end{equation}
since the equation was quadratic in terms of $W(r)$.

One solution corresponds
to scalar particles moving away from the black hole and the
other solution corresponds to particles moving toward the black hole. Imaginary parts of the action can only result from the pole
at the horizon. The probability of crossing
the horizon for outgoing particle is \begin{equation}\label{hj10}
    \text{Prob(out)}\propto\exp(-2\textrm{Im} I)=\exp(-2\textrm{Im}W_+).
\end{equation}
By using the residue theorem, the expression for the quantity $W_+$ is
\begin{equation}\label{11}
    W_+=\frac{2i\pi\omega}{f'(r_{AH})},
\end{equation}
where $r_{AH}$ represents the apparent horizon.

\begin{figure}
  \includegraphics[width=0.9\textwidth]{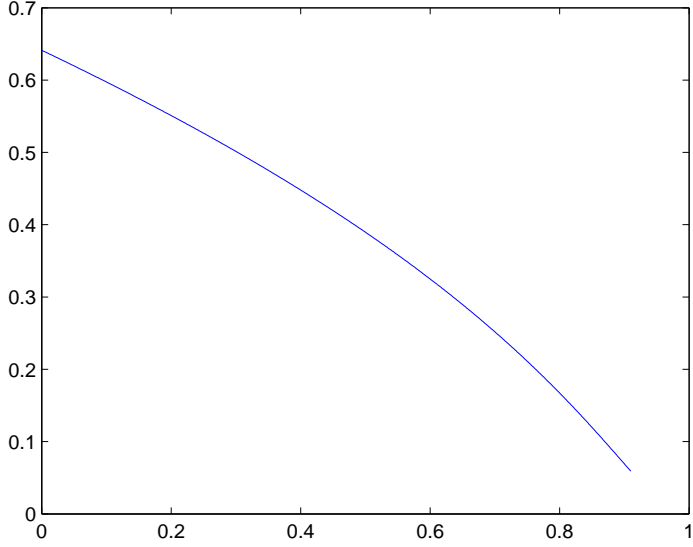}\\
  \caption{Variation of $W_+$ with respect to the ultraviolet correction parameter $\alpha$, for $\Lambda=0$.}\label{figg1}
\end{figure}

When $\Lambda=0$, 
substituting Eqs. (\ref{02}) and (\ref{03}) into Eq. (\ref{11}) yields
\begin{equation}\label{011}
    W_+=\frac{2i\pi(M+\alpha)\omega}{\sqrt{2(M-\alpha)}},
\end{equation}
and the rate of radiation,
\begin{equation}\label{0011}
    \Gamma\propto \exp\left(-\frac{4\pi(M+\alpha)}{\sqrt{2(M-\alpha)}}\omega\right)
,\end{equation}
is plotted in Figure \ref{figg1}.

For $\Lambda\neq0$, the quantity $W_+$ transforms to
\begin{equation}\label{00011}
    W_+=-\frac{3i\pi(4\alpha+r^2_++\sqrt{r^4_++\frac{4}{3}\alpha\Lambda r^4_++16M\alpha})\omega}{\Lambda r_+(r^2_+-r^2_-)}.
\end{equation}

The relation between the outgoing wave and the incoming wave is given by
\begin{equation}\label{roi1}
    \Psi_{out}=\exp\big(-\frac{\pi\omega}{\kappa}\big)\Psi_{in},
\end{equation}
where $\kappa=\big|\frac{f'(r_{AH})}{2}\big|$ is the surface gravity of the black hole and $\Psi_{out}$ and $\Psi_{in}$ are the outgoing wave and the incoming wave, respectively.

The scattering rate of the black hole horizon with a wave function is
\begin{equation}\label{roi2}
    \Big|\frac{\Psi_{out}}{\Psi_{in}}\Big|^2=\exp\big(-\frac{2\pi\omega}{\kappa}\big)=\exp(-2\textrm{Im}W_+),
\end{equation}
since the classical theory of black holes tells us that an incoming particle is absorbed with a probability of one.

\begin{figure}[h!!]
  \includegraphics[width=0.9\textwidth]{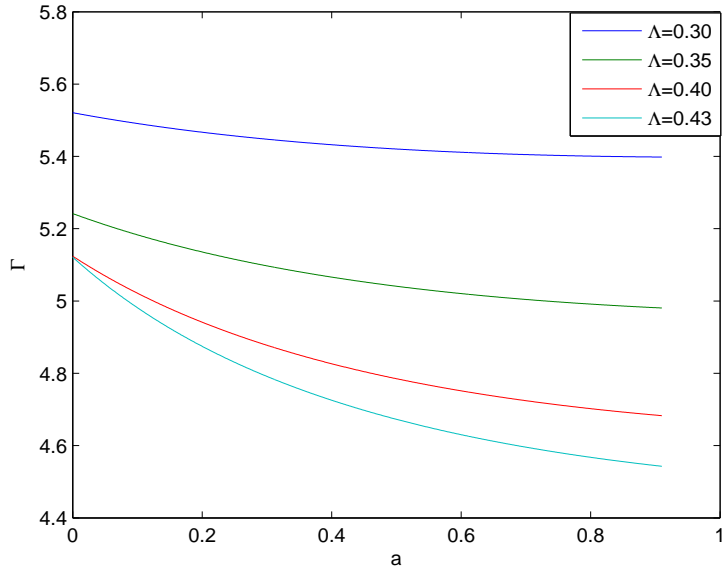}\\
  \caption{Variation of $W_+$ with respect to the ultraviolet correction parameter $\alpha$, for different values of the cosmological constant $\Lambda$.}\label{fig1}
\end{figure}

By substituting into Eq. (\ref{hj10}), the probability $\Gamma$ of the outgoing particle can be expressed as
\begin{equation}\label{hj12}
    \Gamma\propto\exp(-2\textrm{Im}W_+)=\exp\Big(\frac{6\pi(4\alpha+r^2_++\sqrt{r^4_++\frac{4}{3}\alpha\Lambda r^4_++16M\alpha})\omega}{\Lambda r_+(r^2_+-r^2_-)}\Big).
\end{equation}
Its behavior is plotted in Figure \ref{fig1}.

 We can see that the probability $\Gamma$ of the outgoing particle  decreases for an increasing  cosmological constant. This confirms the fact that dark energy reduces the rate of radiation, as demonstrated for the Reissner-Nordstr\"om black hole\cite{mah}.

\begin{figure}[h!]
  \includegraphics[width=0.9\textwidth]{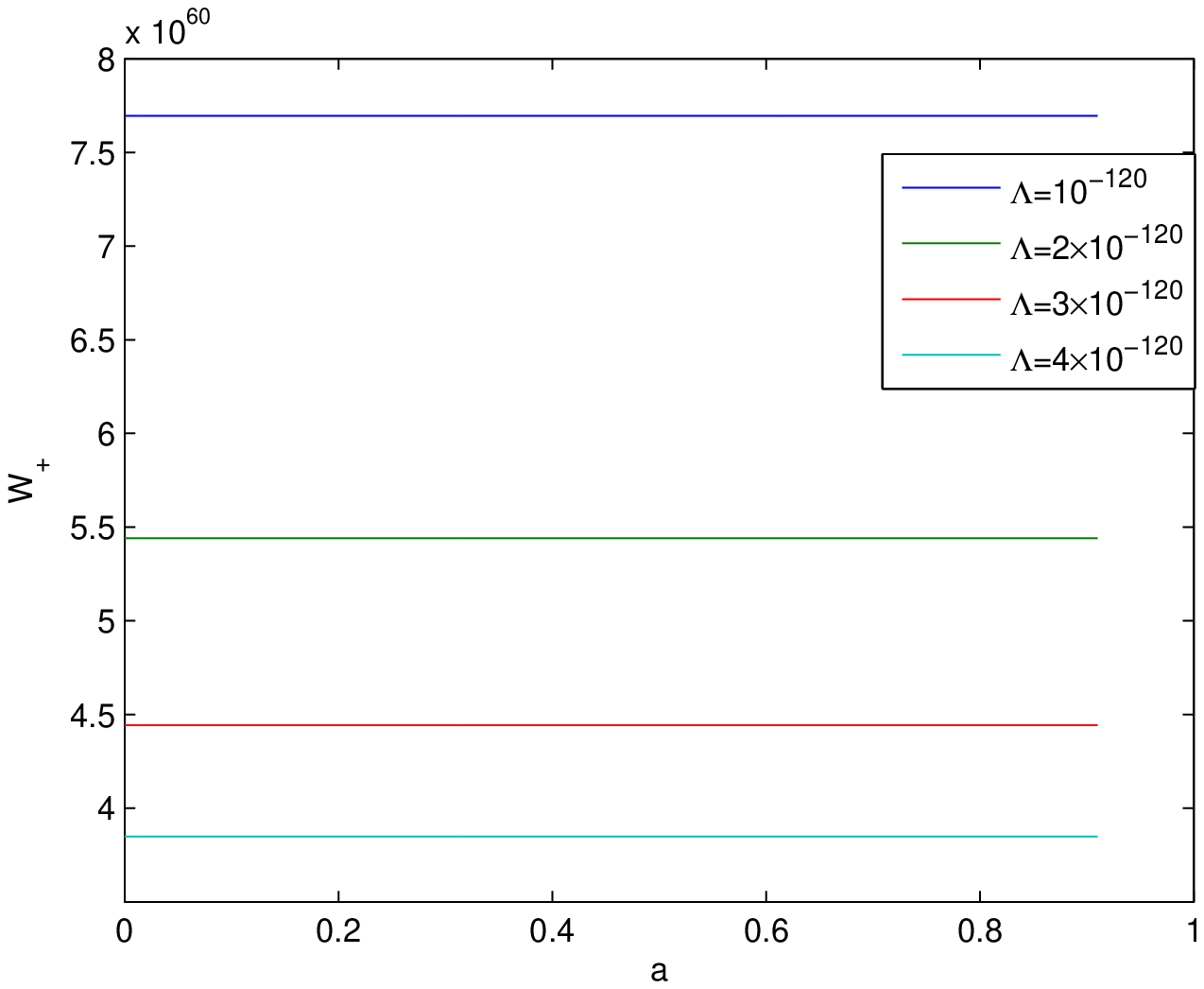}\\
  \caption{Variation of $W_+$ with respect to the ultraviolet correction parameter $\alpha$, for different values of the cosmological constant $\Lambda\sim10^{-120}$.}\label{fig3}
\end{figure}

The actual value of the cosmological constant is slightly less than these values ($\sim$$10^{-120}$)\cite{barrow}. Considering that assertion, we plot the behavior of the quantity $W_+$ with respect to the ultraviolet correction parameter $\alpha$  in Figure \ref{fig3}.
From this figure, we can remark that, for the given value of $\Gamma$, this quantity seems to be independent of the ultraviolet correction parameter $\alpha$, indicating that, when dark energy is considered, the effect of the ultraviolet correction becomes less perceptible.

\section{Conclusion}
\label{conc}

In summary, we have used the Hamilton-Jacobi method to investigate Hawking radiation of a five-dimensional Lovelock black hole. Explicitly, we have plotted the behavior of the rate of radiation from the black hole. Figure \ref{figg1} represents the variation of the rate of radiation with respect to the ultraviolet correction parameter $a$, when $\Lambda=0$, while Figure \ref{fig1} represents the variation of the rate of radiation with respect to the ultraviolet correction parameter $a$, for different values of the cosmological constant $\Lambda\neq0$. We can conclude through these figures that the black hole radiates at a slower rate when  the ultraviolet correction or the cosmological constant are increased. The actual value of the cosmological constant is $\Lambda\sim10^{-120}$ and so the presence of the cosmological constant makes the effect of the ultraviolet correction on the black hole radiation negligible.

\section*{Acknowledgments}
We aim at sending our sincere grateful to the reviewers for their
reports on our paper and their availability. The second author thanks the IHES for hospitalities through the Launsbery Foundation and, the IMU for travel support throughout a grant from the Simons Foundation during the writing of this paper.

\end{document}